\documentclass[12pt]{article}
\usepackage{latexsym,graphicx,color,times}
\usepackage{colordvi}
\usepackage[portrait,margin=1.0in]{geometry}
\def\pd{\partial}

\def\rfig#1{Fig.\ref{fig:#1}}

\long\def\comment#1{}

\newcommand{\be}{\begin{equation}}
\newcommand{\ee}{\end{equation}}

\def\Rdd{\Black}
\def\figdir{./pix}

\def\req#1{(\ref{eq:#1})}
\def\rfig#1{Fig.\ref{fig:#1}}

\def\macname{hank_strauss}
\def\figdira{/Users/{\macname}/Documents/papers/disruption/iter_avde/pix}
\def\figdirag{/Users/{\macname}/Documents/papers/disruption/iter_avde/gnuplot}
\def\figdirb{/Users/{\macname}/Documents/papers/disruption/gnuplot}
\def\figdirg{/Users/{\macname}/Documents/papers/runaways/gnuplot}
\def\figdirr{/Users/{\macname}/Documents/papers/runaways/pix}
\def\figdird3d{/Users/{\macname}/Documents/progs/m3d/d3d-lock}
\def\figditer{/Users/{\macname}/Documents/progs/m3d/itlock}
\def\figdirw{/Users/{\macname}/Documents/papers/disruption/rwtm/gnuplot}
\def\figdircob{/Users/hank_strauss/Documents/progs/m3dc1/JETtime}
\def\figdirnstx{/Users/{\macname}/Documents/progs/m3d/nstx}
\def\figdirm{/Users/{\macname}/Documents/progs/m3d/mst}
\def\figdirp5{/Users/{\macname}/Documents/papers/disruption/rwtm/deltap/paper5}
\def\figdirden{/Users/{\macname}/Documents/papers/disruption/rwtm/mst-density}
\def\figdirdcaf{/Users/{\macname}/Documents/papers/disruption/decaf}
\def\figdirnstx{/Users/{\macname}/Documents/progs/m3d/nstx}
\def\figdirntm{/Users/{\macname}/Documents/papers/disruption/ntm}
\def\figdirkstr{/Users/{\macname}/Documents/papers/disruption/rwtm/kstar}
\def\figdirliq{/Users/{\macname}/Documents/papers/disruption/rwtm/liq95}
\def\figdirdd{/Users/{\macname}/Documents/papers/disruption/rwtm/d3d}
\def\figdir{./pix}

\begin{document}
\begin{center}
{\Large\bf Eliminating  Tokamak Disruptions with Feedback} \\
{\large H. R. Strauss} \\
{HRS Fusion, West Orange NJ 05072 USA} \\
{email: hank@hrsfusion.com} \\
\end{center}

\long\def\comment#1{}
\def\macname{hank_strauss}
\def\figdira{/Users/{\macname}/Documents/papers/disruption/iter_avde/pix}
\def\figdirag{/Users/{\macname}/Documents/papers/disruption/iter_avde/gnuplot}
\def\figdirb{/Users/{\macname}/Documents/papers/disruption/gnuplot}
\def\figdirg{/Users/{\macname}/Documents/papers/runaways/gnuplot}
\def\figdirr{/Users/{\macname}/Documents/papers/runaways/pix}
\def\figdird3d{/Users/{\macname}/Documents/progs/m3d/d3d-lock}
\def\figditer{/Users/{\macname}/Documents/progs/m3d/itlock}
\def\figdirw{/Users/{\macname}/Documents/papers/disruption/rwtm/gnuplot}
\def\figdircob{/Users/hank_strauss/Documents/progs/m3dc1/JETtime2010}
\def\figdirnstx{/Users/{\macname}/Documents/progs/m3d/nstx}
\def\figdirm{/Users/{\macname}/Documents/progs/m3d/mst}
\def\figdirp5{/Users/{\macname}/Documents/papers/disruption/rwtm/deltap/paper5}
\def\figdirden{/Users/{\macname}/Documents/papers/disruption/rwtm/mst-density}
\def\figdirdcaf{/Users/{\macname}/Documents/papers/disruption/decaf}
\def\figdirnstx{/Users/{\macname}/Documents/progs/m3d/nstx}
\def\figdirntm{/Users/{\macname}/Documents/papers/disruption/ntm}
\def\figdirkstr{/Users/{\macname}/Documents/papers/disruption/rwtm/kstar}
\def\figdirliq{/Users/{\macname}/Documents/papers/disruption/rwtm/liq95}
\def\figdirdd{/Users/{\macname}/Documents/papers/disruption/rwtm/d3d}
\def\figdir{.}

%\begin{document}

%\title{\hspace{10cm} {\bf T-S} \\
%\title{\rightline{{\bf T-S}} \\

%  \centerline{\Red{Recent Results}}
\section*{\large{\bf Abstract}}
%{\fontsize{9pt}{12pt}\selectfont\hspace{1cm}
{\small 
Many disruptions are caused by resistive wall tearing modes (RWTM). A database
of DIII-D locked mode disruptions provides two main disruption criteria, which are
shown to be signatures of RWTMs. The first is that the q = 2 rational surface must be
sufficiently close the resistive wall surrounding the plasma to interact with it.
If active feedback is used, this implies that RWTMs can be prevented from causing
major disruptions. This is demonstrated in simulations.
The second criterion is that the current profile is sufficiently
peaked.  This is caused by edge cooling, such as by  impurity radiation
and turbulence, which suppress edge current and temperature. This implies the
disruptions are not caused by neoclassical tearing modes (NTM), because the bootstrap
current is also suppressed. 
The dependence of the critical internal inductance  on elongation is given, which suggests that elongation might 
be used as an actuator to prevent disruptions.
At high $\beta,$ resistive wall modes (RWM)  can be stabilized
with feedback. Feedback also stabilizes high $\beta$   RWTMs, as shown in NSTX data 
and in simulations.  These results
suggest that RWTM disruptions in ITER might be prevented using the resonant magnetic
perturbation (RMP) coils.}
%\fontsize{12pt}

%\begin{section}{\large{\bf Introduction}}
  \section*{\Rdd{\bf 1.  Introduction}}
 % {\small \section{\bf Introduction}}
Disruptions have been considered to be a major obstacle to tokamak fusion.
Many disruptions are caused by resistive wall tearing modes (RWTM) \cite{jet21,iter21,d3d22,mst23,
%gimblett,finn95,betti}. 
finn95,betti}. 
They can be prevented with feedback. 

There are two main criteria \cite{sweeney2017}  for locked mode disruptions in DIII-D, which will
 be shown to be signatures of RWTMs.
The first and most important is that $\rho_{q2}$ exceeds a threshold, which in 
DIII-D is 
  $\rho_{q2} > 0.75.$ 
   Here $\rho_{q2}$ is the $q = 2$ radius %$R_{q2} - R_0$ 
    normalized to wall radius.
This shows that the disruption is caused by a tearing mode
   close enough to the wall to interact with it. 
This is the main characteristic of RWTMs, which
 makes feedback active control
   possible.
The criterion depends on the distance of the wall from the plasma.
     
A second criterion is that the internal inductance  exceeds a critical value,
  which depends on the geometry of the equilibrium. It can be expressed as
a threshold $l_i / q_{95},$ which for DIII-D is  
   $l_i/q_{95} > 0.28.$  The current must be sufficiently peaked. This can be
   caused by edge cooling and other precursors. Edge cooling suppresses bootstrap current, 
    so the disruptions are not caused by NTMs. The dependence 
     of the critical internal inductance on elongation is given, 
suggesting that it might be an actuator to prevent disruptions. 

% RWTMs are low $\beta,q$ but  similar to high $beta,q$  RWMs, RPRWM -
Nonlinear simulations demonstrate that
   RWTMs can grow to much larger amplitude than ideal wall tearing modes (TM), 
and cause a complete thermal quench. In contrast, tearing modes with an ideal
wall boundary saturate at low amplitude, which can cause a minor disruption.
Feedback can emulate an  ideal wall, which can prevent major disruptions.

RWTMs are also found at high $\beta,$ in NSTX \cite{sabbagh2010,nstx25} and KSTAR \cite{kstar}. 
Feedback control of
resistive wall modes (RWM) also controls RWTMs. This is seen in NSTX data, as well as
in simulations.

 ITER resonant magnetic perturbation (RMP) coils
 might be used for feedback.
Prevention of disruptions using feedback could greatly improve the prospects for fusion.

%\end{section}
%\begin{section}{\Rdd {\bf  DIII-D database and $\rho_{q2}$, $l_i / q_{95}$ disruption  criteria}}
\section*{\Rdd {\bf 2. DIII-D database and $\rho_{q2}$, $l_i / q_{95}$ disruption  criteria}}
\begin{figure}[h]
 \begin{center}
\includegraphics[height=4.5cm] {\figdir/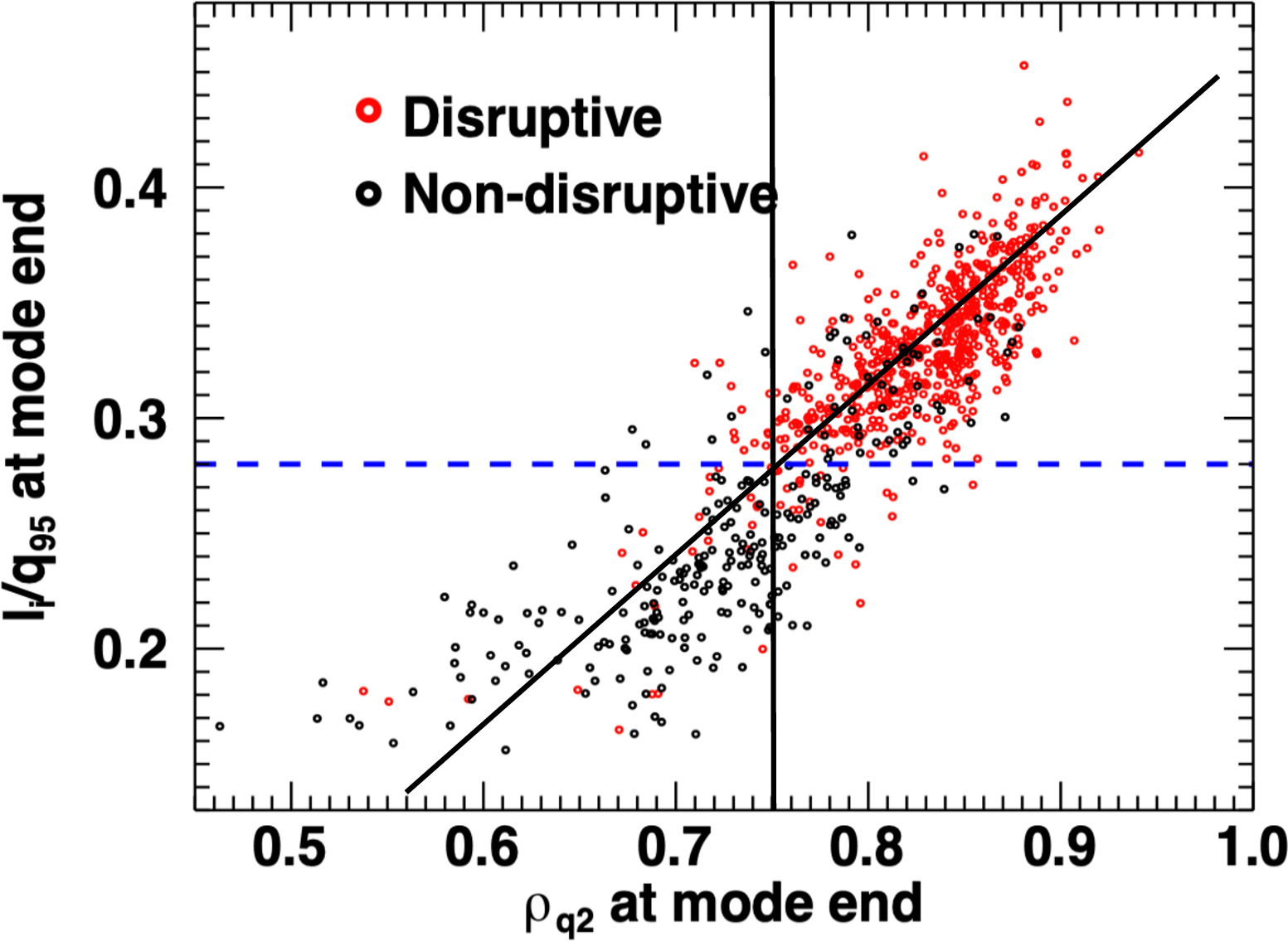}
\end{center}
\caption{\it
  Disruptivity in a DIII-D locked mode disruption database,
adapted from \cite{sweeney2017} with IAEA permission.} \label{fig:sweeney}
\end{figure}

Sheared rotation stabilizes tearing modes. When the rotation locks, tearing modes can become
unstable. 
Disruption precursors cause edge cooling, which causes the current to contract. This
can cause two main disruption criteria to be satisfied.
This is shown in a database \cite{sweeney2017} of DIII-D locked mode disruptions.
\rfig{sweeney} shows disruptivity in the $(\rho_{q2},l_i/q_{95})$ plane. The
horizontal coordinate $\rho_{q2}$ is the minor radius of the $q = 2$ rational
surface, normalized to the plasma radius.
The  disruption criterion  is $\rho_{q2} \ge .75$. 
The vertical coordinate is the internal inductance divided 
by $q_{95}.$
The sloping line is a fit \cite{sweeney2017} to the data, 
\be l_i / q_{95} = (2/3) \rho_{q2} - .22 \label{eq:fit} \ee
This couples the conditions on $l_i / q_{95}$ and  $\rho_{q2}.$
The  critical $l_i/ q_{95} = 0.28$, which is a measure of contraction 
of the current profile. A peaked profile has higher $l_i$ than a broad profile.
The $l_i / q_{95}$ criterion will be discussed in Section 6. %\rsec{current}.

These criteria  imply that the disruptions are caused by RWTMs.
The modes are tearing, because $\rho_{q2} < 1,$ and resistive wall modes, because $\rho_{q2}
> 0.75,$ allowing wall interaction and making feedback possible.
The criterion is the condition for a lobe of the  mode to reach the wall, 
so the mode ``knows" the wall boundary condition.
%RWTMs grow to large amplitude, sufficient for a complete thermal quench.
%If the wall is ideal, the modes only cause minor disruptions.
%Feedback can emulate an ideal wall and prevent major disruptions.
%\end{section}
%\begin{section}{\Rdd{\bf  $\rho_{q2}$ criterion}}
\begin{figure}[h] 
\begin{center}
\includegraphics[height=4.2cm]{\figdir/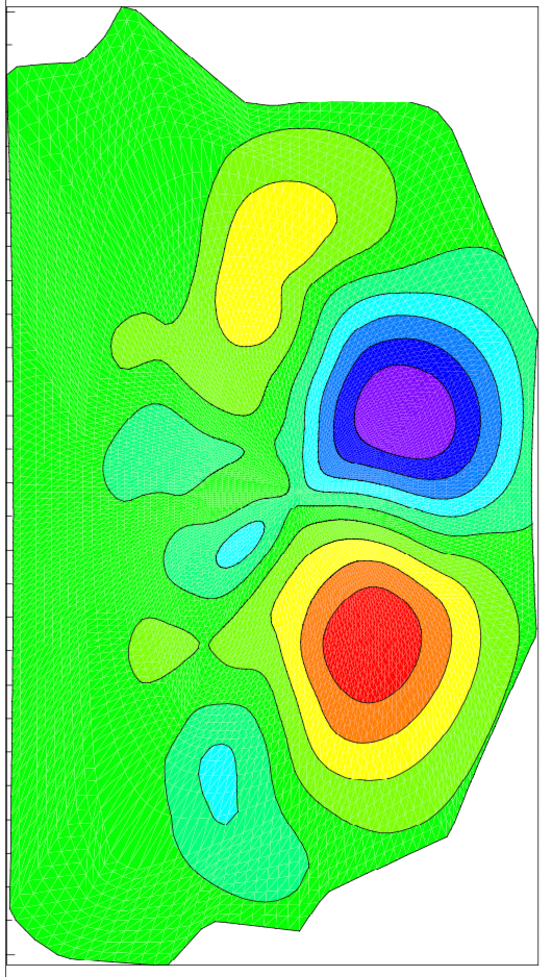}(a)
\includegraphics[height=3.0cm]{\figdir/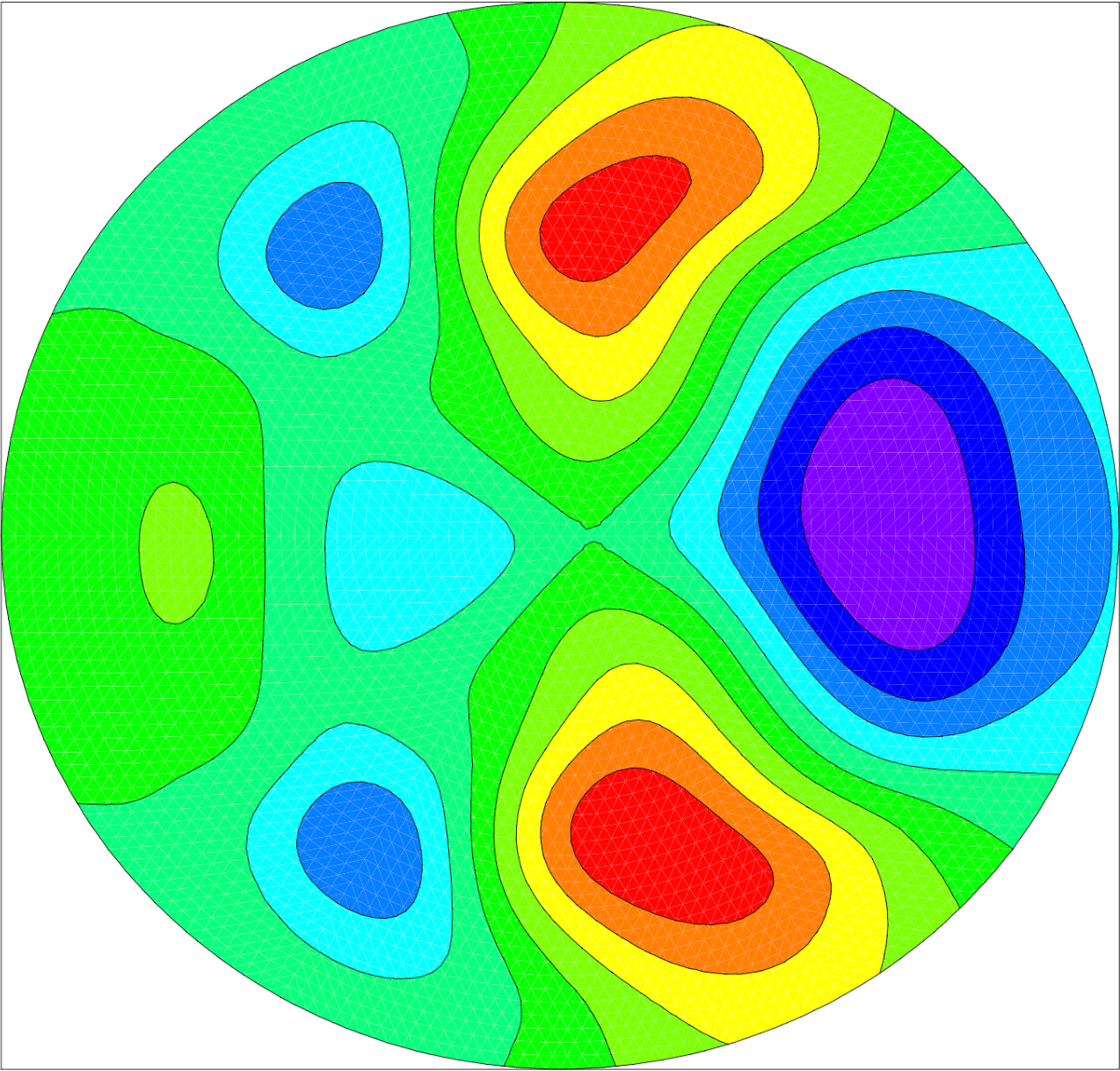}(b)
\includegraphics[height=4.2cm]{\figdir/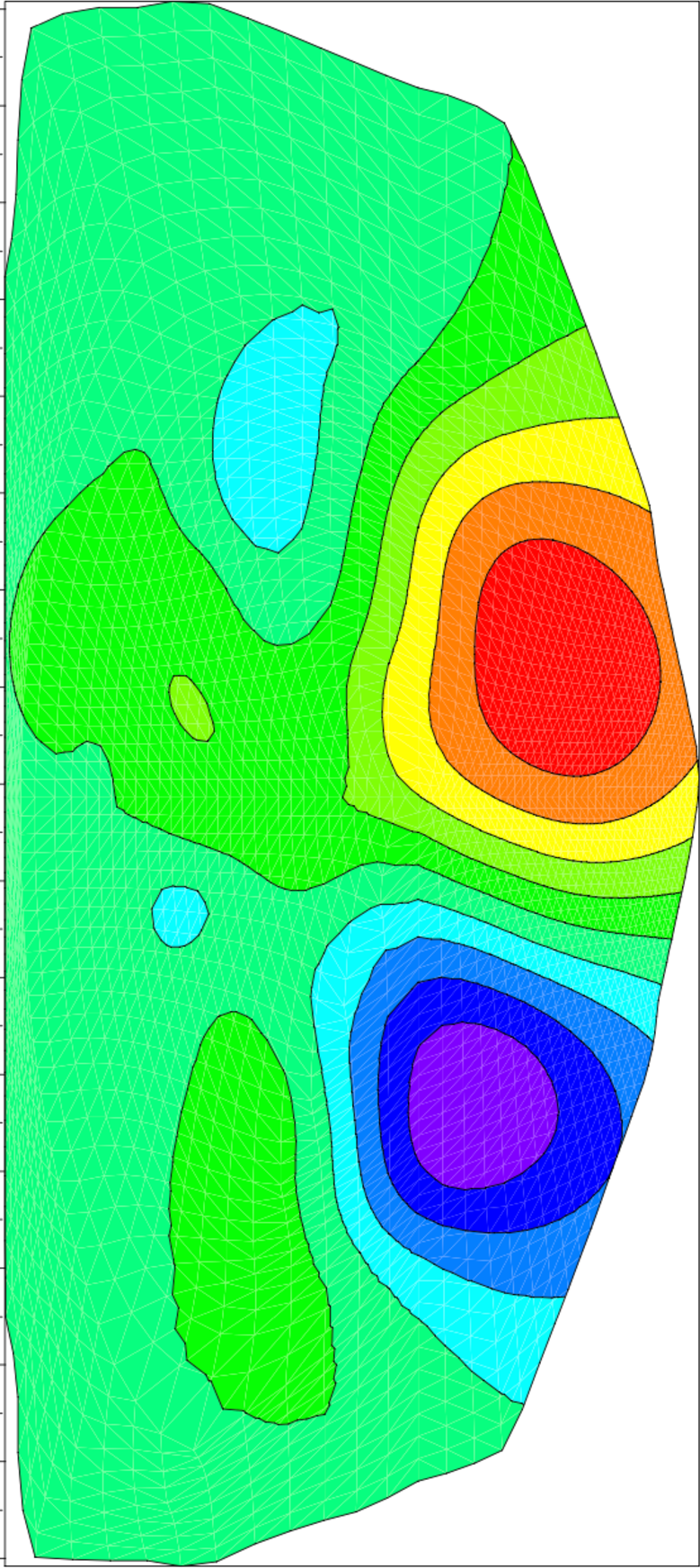}(c)
\includegraphics[height=4.2cm]{\figdir/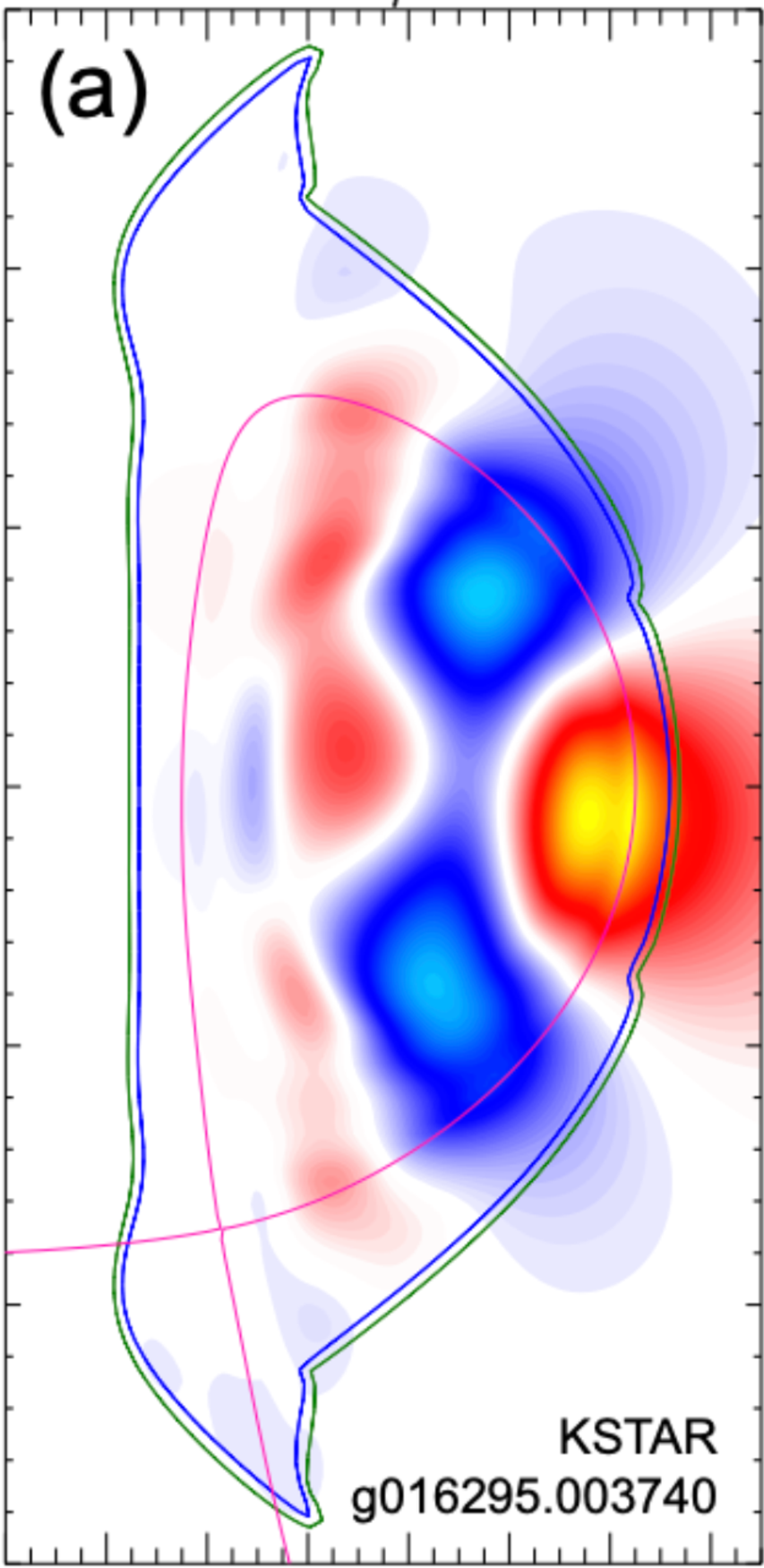}(d)
\end{center}
\caption{\it
(a) Perturbed $\psi$ in DIII-D simulations, reproduced from \cite{d3d22} with AIP permission,
(b) MST - based  sequence in  Section 4, reproduced from \cite{model} with AIP permission; (c) NSTX 
reproduced from \cite{nstx25} with IAEA permission, (d) KSTAR adapted from \cite{kstar}
with IAEA permission.}
\label{fig:lobe}
\end{figure}
\rfig{lobe} gives examples of simulations of RWTMs in several experiments: 
DIII-D \cite{d3d22}, Madison Symmetric Torus (MST) \cite{mst23,model}, NSTX \cite{sabbagh2010,nstx25} , and KSTAR
\cite{kstar}.
Shown is the perturbed magnetic flux $\psi$ in  poloidal plane. 
The normalized distance to the wall is $\rho_w = \rho_{q2} + 1/ (k_\perp a),$
where the wavenumber is the normalized lobe radius $(k_\perp a \approx m / \rho_{q2}.$
For mode number $m = 2$ 
\be  \rho_{q2} \approx \frac{\rho_w}{1 + 1/2} \approx  0.8 \label{eq:rhoq2} \ee
taking  $\rho_w \approx 1.2$ in examples (a) - (d).
More accurately \cite{nstx25, model,nf-iaea24},
%$\rho_{q2} \ge  0.75$ for $\rho_w = 1.2.$ %The minimum $\rho_{q2}$ occurs for $\rho_w = 1,$
%The  more accurate calculation \cite{model,nf-iaea24}, 
%as described briefly  in Section 6, 
the onset condition is
 $\rho_{q2} = 0.625 \rho_{w},$ giving $\rho_{q2} = 0.75$ for $\rho_w = 1.2.$
The maximum wall distance for a RWTM, from \req{rhoq2}, for $\rho_{q2} < 1,$ is  $\rho_w = 1.5.$ 
Otherwise it is 
a no wall tearing mode 
\cite{nstx25}.

The importance of the $\rho_{q2}$ criterion is that the tearing mode is a RWTM. The mode
behavior depends on the  wall boundary condition.
RWTMs grow to large amplitude, sufficient for a complete thermal quench.
If the wall is ideal, the modes only cause minor disruptions.
Feedback can emulate an ideal wall and prevent major disruptions.

It should be pointed out  that according to \rfig{sweeney}, not all disruptions satisfy the
two criteria, and not all shots that satisfy the criteria disrupt. These cases should
be investigated. It is interesting that all the cases align with the fit \req{fit}.
%\end{section}

%\begin{section}{\large{\bf  Feedback stabilization of RWTM}}
 \section*{\large{\bf 3.  Feedback stabilization of RWTM}}
Feedback experiments on DIII-D and RFX \cite{hanson,piovesan} showed stabilization of
RWMs with $q_a = 2.$ Feedback was used to stabilize high $\beta$ RWM in
DIII-D \cite{okabayashi2009}, 
NSTX \cite{sabbagh2010},
and KSTAR \cite{kstar}. %Complex feedback in DIII-D \cite{okabayashi2} 
%prevented mode locking.
An attempt was made to control tearing modes \cite{zanca2} 
with $q_a > 2$, but encountered density limit disruptions.
%Saddle coils which sense normal magnetic perturbations $b_n$ % \propto \pd \psi / \pd l$ %,
%and probes which sense tangential  $b_l \propto \pd \psi / \pd n$ % are used,
%are modeled,
%which is fed back into the evolution of  $\psi$ at the wall.
%The measurement can be outside the wall, by continuity of $b_n.$
%Saddle coil sensors were used in RFX [Zanca (2012)] filtering the aliasing error.
%Probes were used in DIII-D [Hanson 2014].
The feedback is done in the simulations with saddle coils placed outside the
resistive wall. Sensors inside the wall detect the mode magnetic perturbations.

In simulations,  feedback is added %  normal  gain  $\gamma_w$ and rotation $\Omega_w$
to the thin wall boundary condition,
\be \frac{\pd \psi_w}{\pd t} = \frac{r_w}{\tau_{wall}} ( \psi'_{vac} -  \psi_p' ) %+\psi'_{fbk} )
 - \gamma_w \psi_{sensor}\Psi_{coil} 
\label{eq:thin1} \ee
%\be \frac{r_w}{\tau_{wall}} \psi'_{fdbk} =  - \gamma_w \psi_{sensor}\Psi_{coil} 
%\ee
where $\psi_w$ is the magnetic flux at the wall, 
$\psi'_{vac}$ is the vacuum magnetic flux normal derivative at the wall excluding the   
contribution of the feedback coils, $\psi_p'$ is the magnetic flux normal derivative
from the plasma at the wall,
$\gamma_s$ is the  gain, $\psi_{sensor}$ is $\psi$ at the  sensors, and $\Psi_{coil}$ is the
normalized $\psi$ of the  coils on the wall. This is more realistic than the  simplified  feedback model used in 
\cite{nstx25,nf-iaea24}. Examples of $\Psi_{coil}$ are shown in \rfig{mstfeed}(b) 
and \rfig{nstxfeed}(a).

% \end{section}

%\begin{section}{\Rdd{\bf  MST extended equilibria and feedback simulations }} \label{sec:mst}
\section*{\Rdd{\bf 4.  MST extended equilibria and feedback simulations }} \label{sec:mst}
\begin{figure}[h] 
\begin{center}
\includegraphics[height=5.0cm] {\figdir/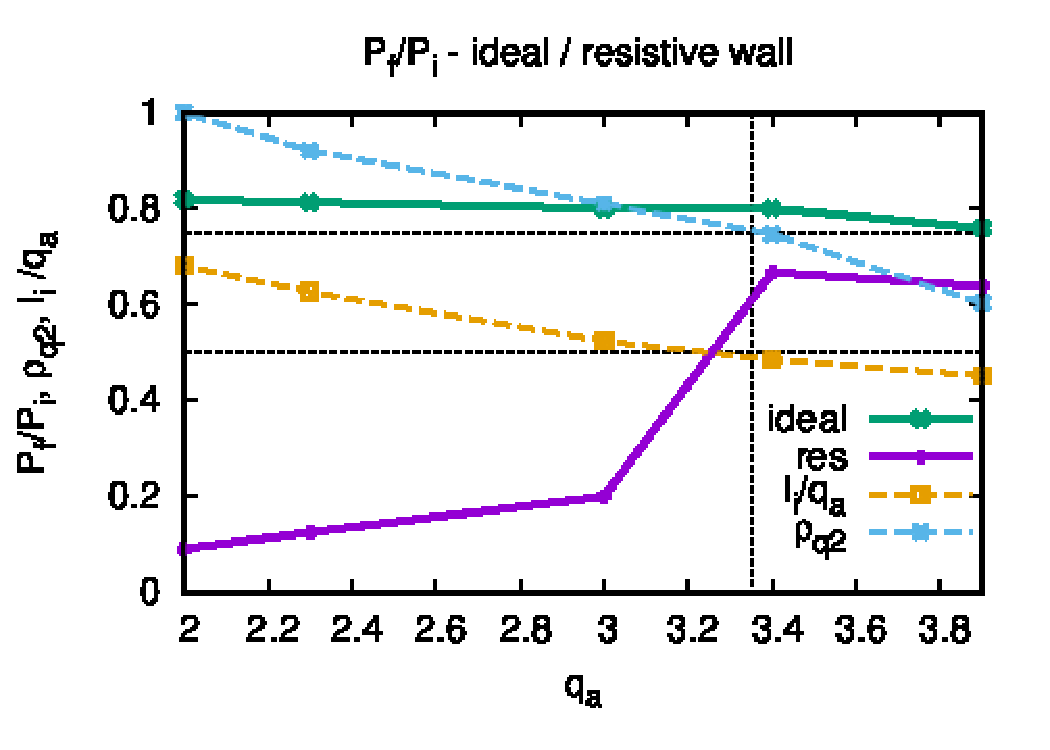}
\end{center} \caption{ \it A sequence of MST  equilibria with $2.3 \le q_a \le 3.9.$
 Total pressure drop  $P_{final}/P_{initial}$ for ideal and resistive wall, $\rho_{q2}$
and $l_i / q_a$ 
as functions of $q_a.$ Major disruptions occur for $\rho_{q2} \ge 0.75,$
$l_i / q_a > 0.5.$
}
\label{fig:mst}
\end{figure}
MST has low $q$, but  does not have disruptions because the wall is ideal on the experimental
  shot timescale. 
In  M3D
\cite{m3d} 
simulations of a sequence of equilibria \cite{model,nf-iaea24}  with $2.3 \le q_a \le 3.9,$ the
resistive wall time was  made artificially short. % Major disruptions occur for resistive wall and
The wall distance was increased to $\rho_w  = 1.2,$ like DIII-D.
The simulations are summarized in \rfig{mst}, 
%For an ideal wall, only minor disruptions occur.
%If an ideal wall can be emulated with feedback, major disruptions
%will be suppressed.
which shows the  total pressure drop  $P_{final}/P_{initial}$ for ideal and resistive wall, 
 $\rho_{q2}$, and $l_i / q_a$
as functions of $q_a.$ 
Here $P_{final}$ is the total pressure after MHD activity, $P_{initial}$ is the initial
total pressure. 
With a resistive wall, major disruptions occur  for $\rho_{q2} \ge 0.75,$
$l_i /q_a > 0.5.$
This is in agreement with the DIII-D database of Section 2, taking into account a  geometry dependent
rescaling \req{scale} of $l_i /q_a.$ The critical $q_a \approx 3.35.$
With an ideal wall, only minor disruptions occur. 

This implies that if feedback is applied to the resistive wall, an ideal wall can be
emulated.
%This is demonstrated in \rfig{mstfeed},
%(b) Pressure $p$  contours  in nonlinear simulation of the $q_a = 3$ case for
% resistive wall,
%(c) perturbed  magnetic flux $\psi$ corresponding to  case (b),
%(c) Pressure $p$  contours  in nonlinear simulation of the $q_a = 3$ case for
%ideal wall.
%(c) perturbed  magnetic flux $\psi$ from coils
%(d) pressure contours with feedback stabilization using the coils.
\begin{figure}[h]
\begin{center}
\includegraphics[height=3.0cm]{\figdir/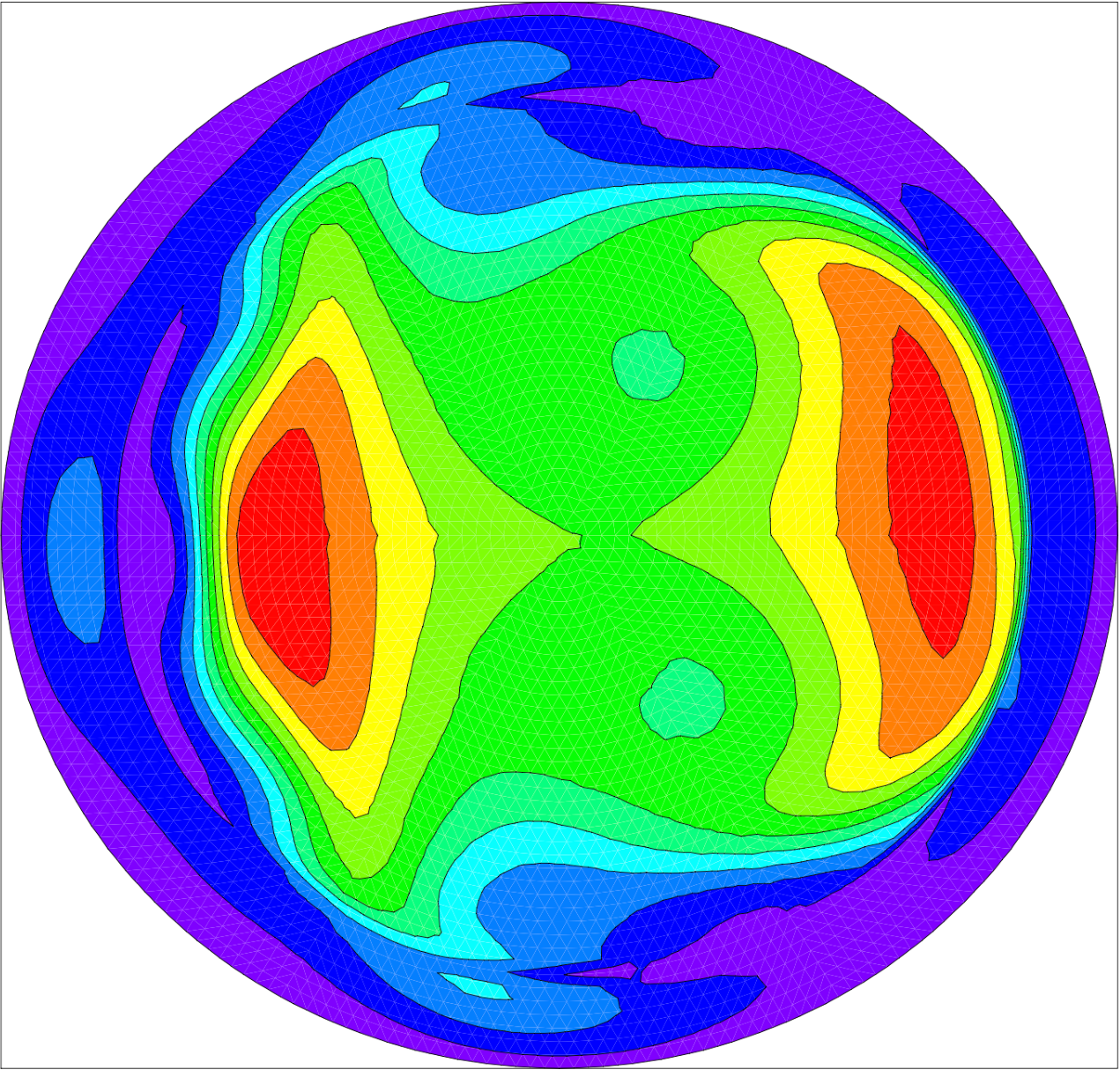}(a)
\includegraphics[height=3.0cm]{\figdir/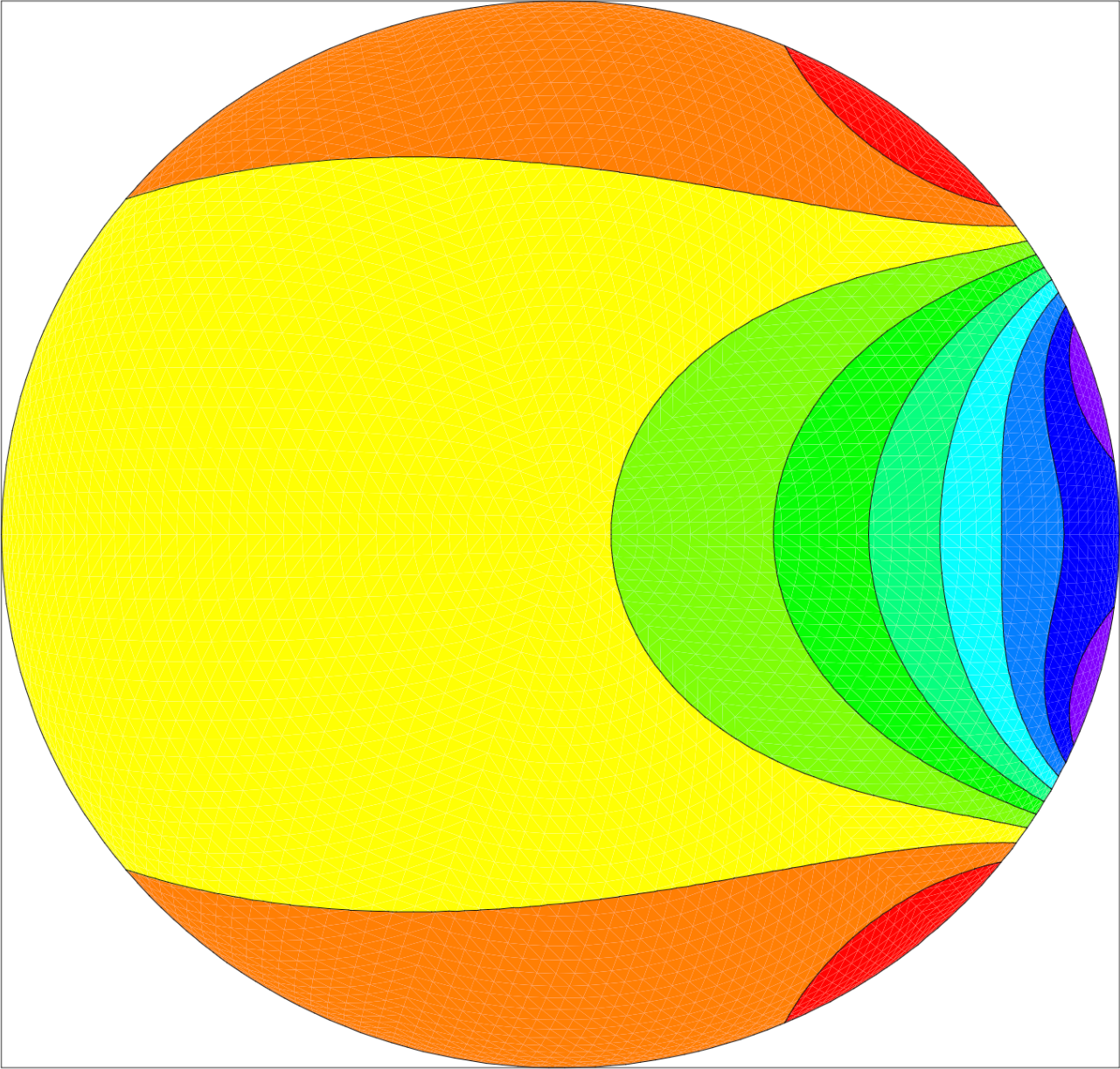}(b)
\includegraphics[height=3.0cm]{\figdir/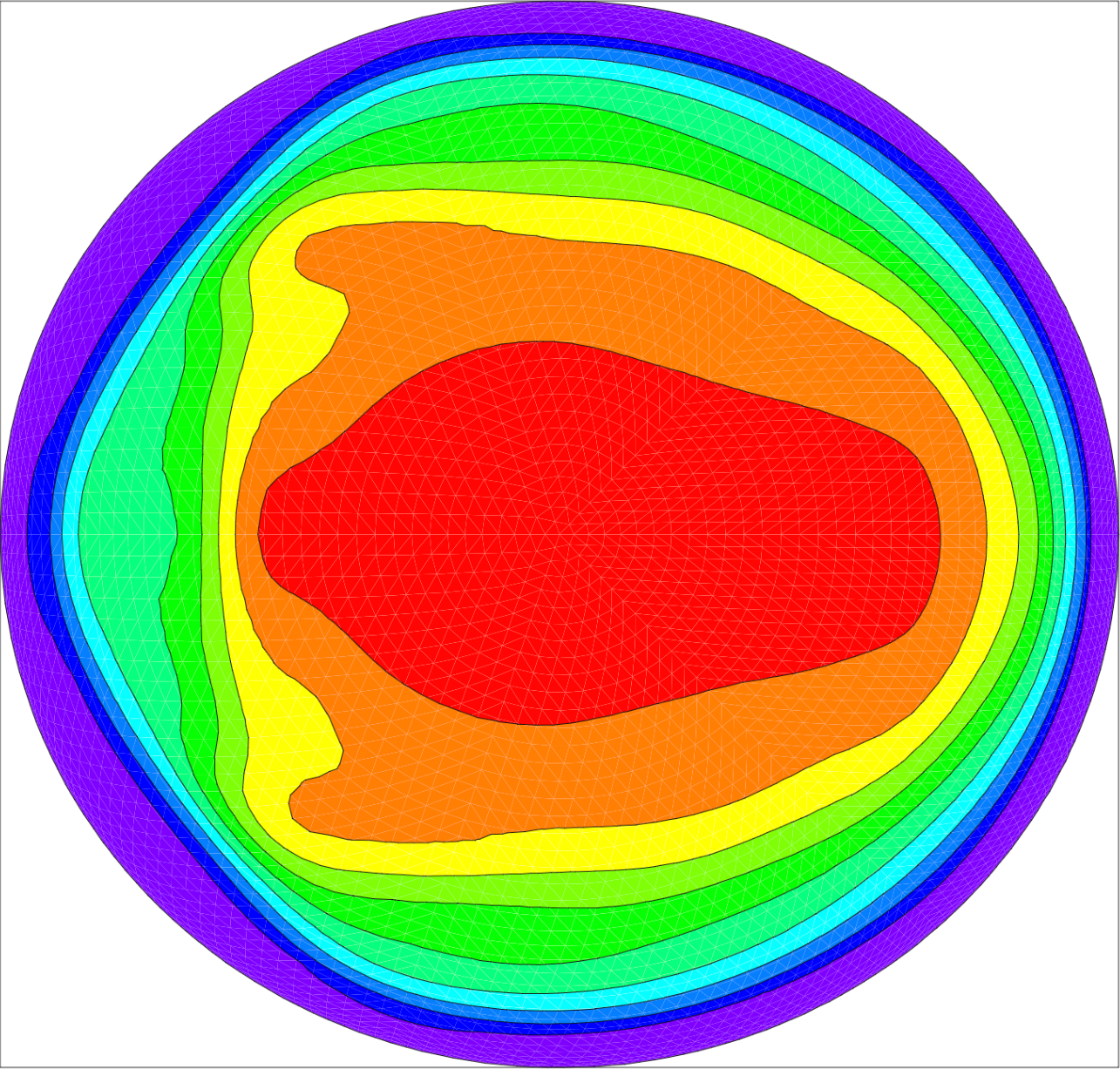}(c)
\end{center}
\caption{\it
(a) pressure contours with $q_a = 3$, resistive wall,  without feedback, in a 
major disruption, reproduced from \cite{model} with AIP permission; 
(b) perturbed  magnetic flux $\psi$ from coils, compare with \rfig{lobe} (b);
%(b) perturbed $\psi$ with resistive wall, $q_a  = 3;$
(c) pressure contours with feedback stabilization using the coils, limiting the
mode to much smaller amplitude.}
\label{fig:mstfeed}
\end{figure}
\rfig{mstfeed} (a) shows a major disruption for $q_a = 3$ and  resistive wall.
With no  feedback,
the  pressure contours show  a very large $(m,n) = (2,1)$ perturbation. 
The accompanying 
 $\psi$ perturbations are  shown in \rfig{lobe} (b).
Feedback is applied using 
magnetic flux $\psi$ from the coils, shown in 
\rfig{mstfeed} (b). 
Comparing  with \rfig{lobe} (b) indicates that the feedback flux $\psi$  roughly
matches $\psi$ of  the perturbation at the wall.
\rfig{mstfeed} (c) shows  pressure contours with feedback stabilization using the coils, limiting the
mode to much smaller amplitude.
% \end{section}
%\begin{section}{\large{\bf NSTX RWTM} }
\section*{\large{\bf 5. NSTX RWTM} }
\begin{figure}[h]
\begin{center}
\includegraphics[width=7.0cm]{\figdir/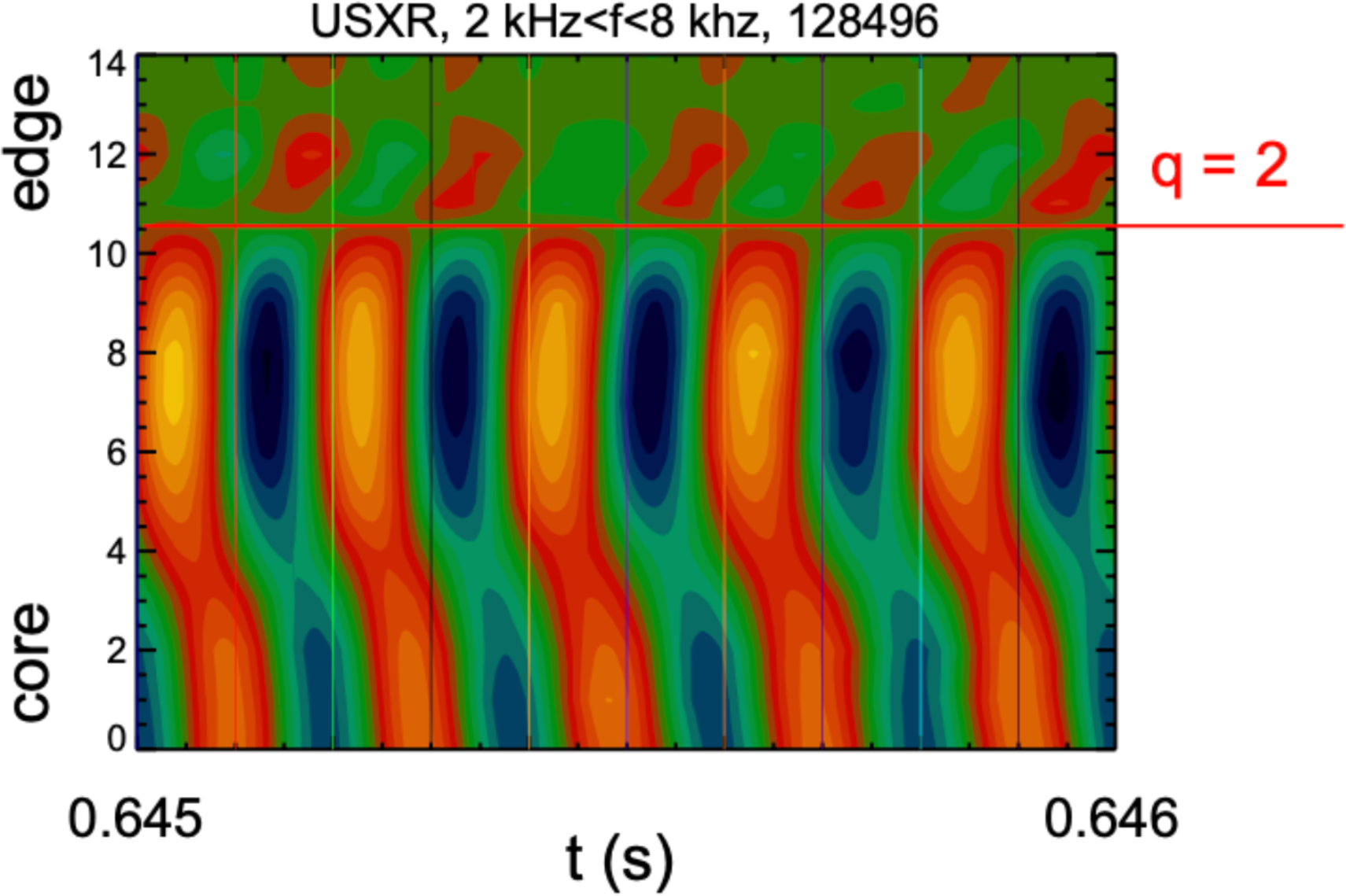} %(b)
\end{center}
\caption{\it
NSTX example  with $\beta_N > 4,$ above the no wall limit.
showing soft X ray emission as a function of radius and time, reproduced from \cite{sabbagh2010}
with IAEA permission.}
\label{fig:nstx}
\end{figure}

Feedback stabilization was used successfully in NSTX to control RWMs. As a kind of
by - product, RWTMs were also controlled. Evidently coils suitable for $(3,1)$ RWMs
also worked for $(2,1)$ RWTMs.
\rfig{nstx} gives an 
NSTX example  with $\beta_N > 4,$ above the no wall limit.
Plotted is soft X ray emission as a function of radius and time.
The soft X ray amplitude of the vertical axis shows
the radial mode structure of
a feedback stabilized  $(2,1)$ mode. The horizontal axis shows the rotation of the
signal. 
It  can be identified as a RWTM  by
its  phase inversion at  $\rho_{q2} = 0.75,$ and because it is feedback stabilized
\cite{sabbagh2010}.
\begin{figure}[h]
\begin{center}
\includegraphics[height=5.6cm] {\figdir/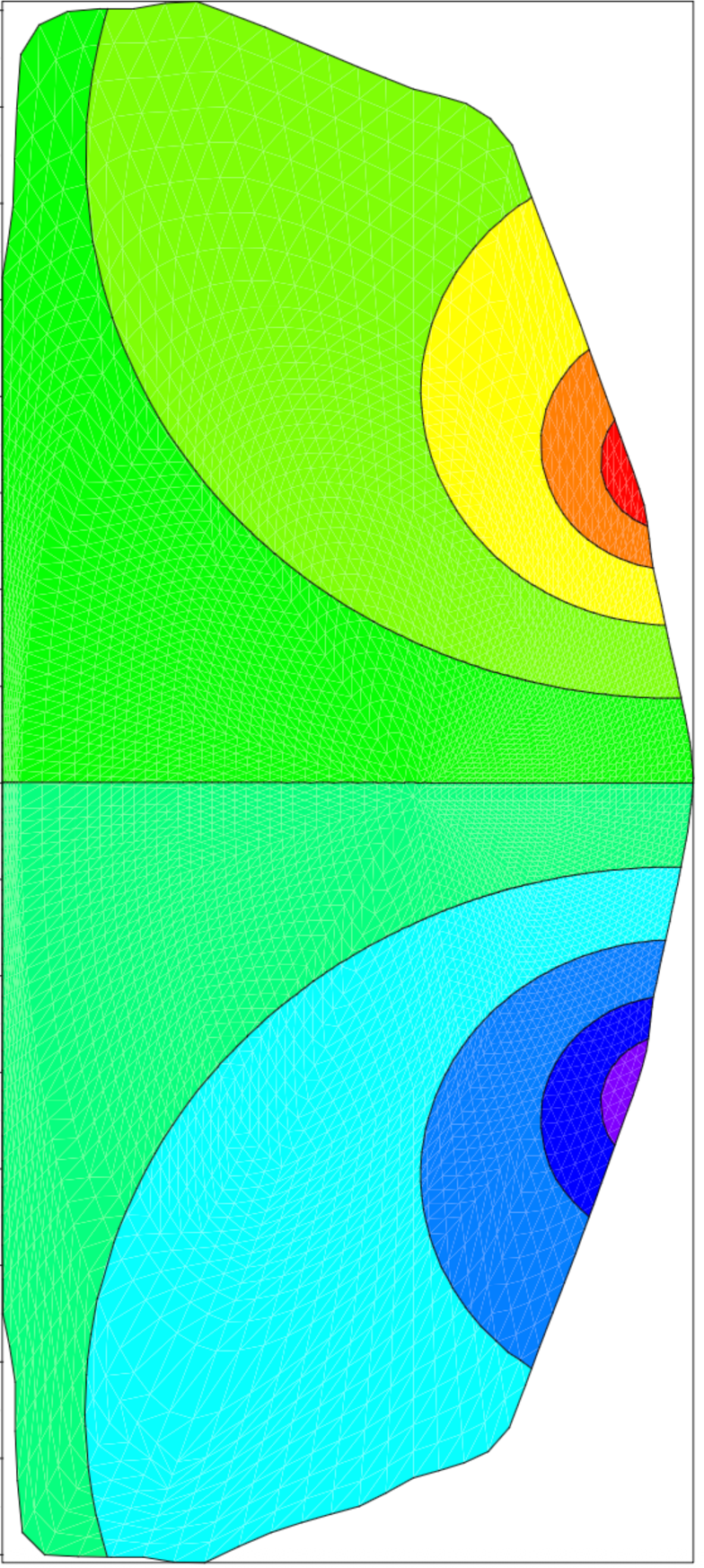}(a)
\includegraphics[height=5.6cm] {\figdir/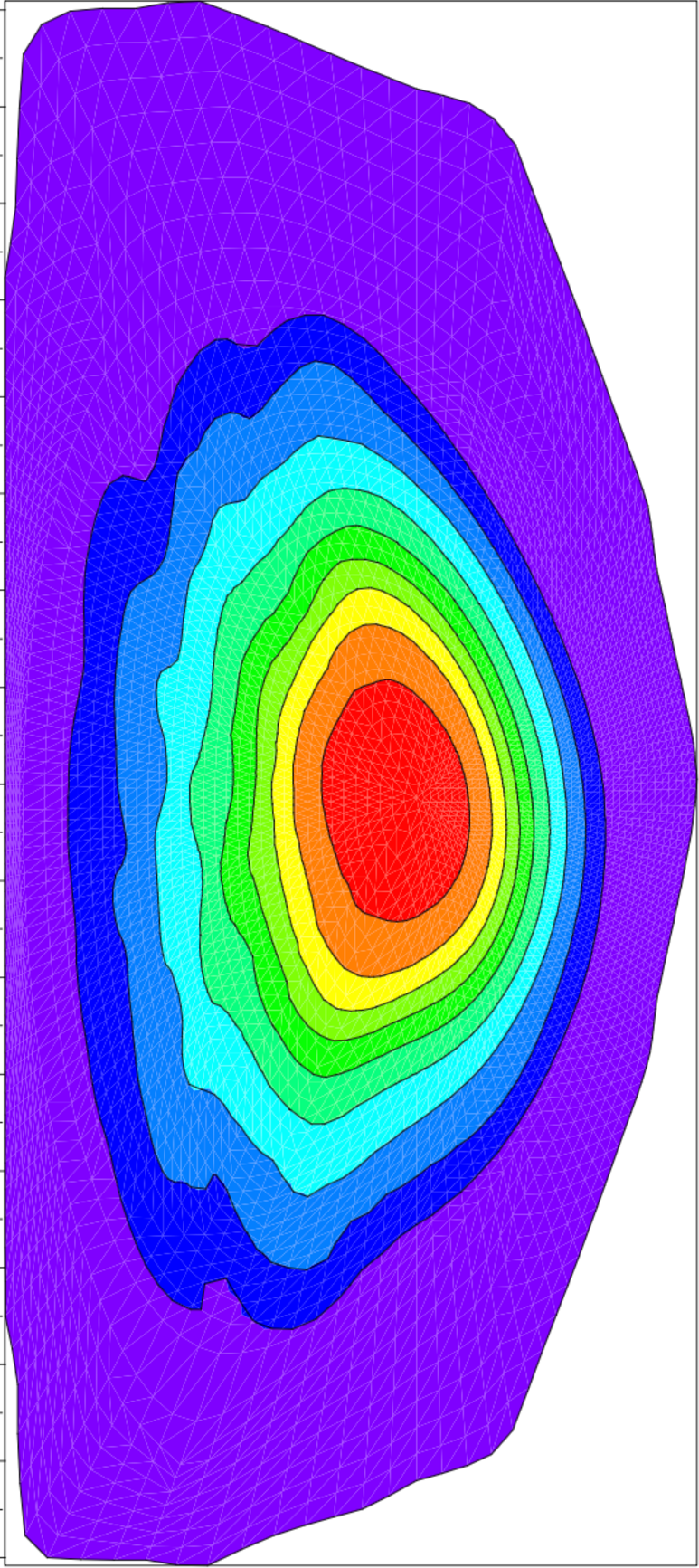}(b)
\includegraphics[height=5.6cm]{\figdir/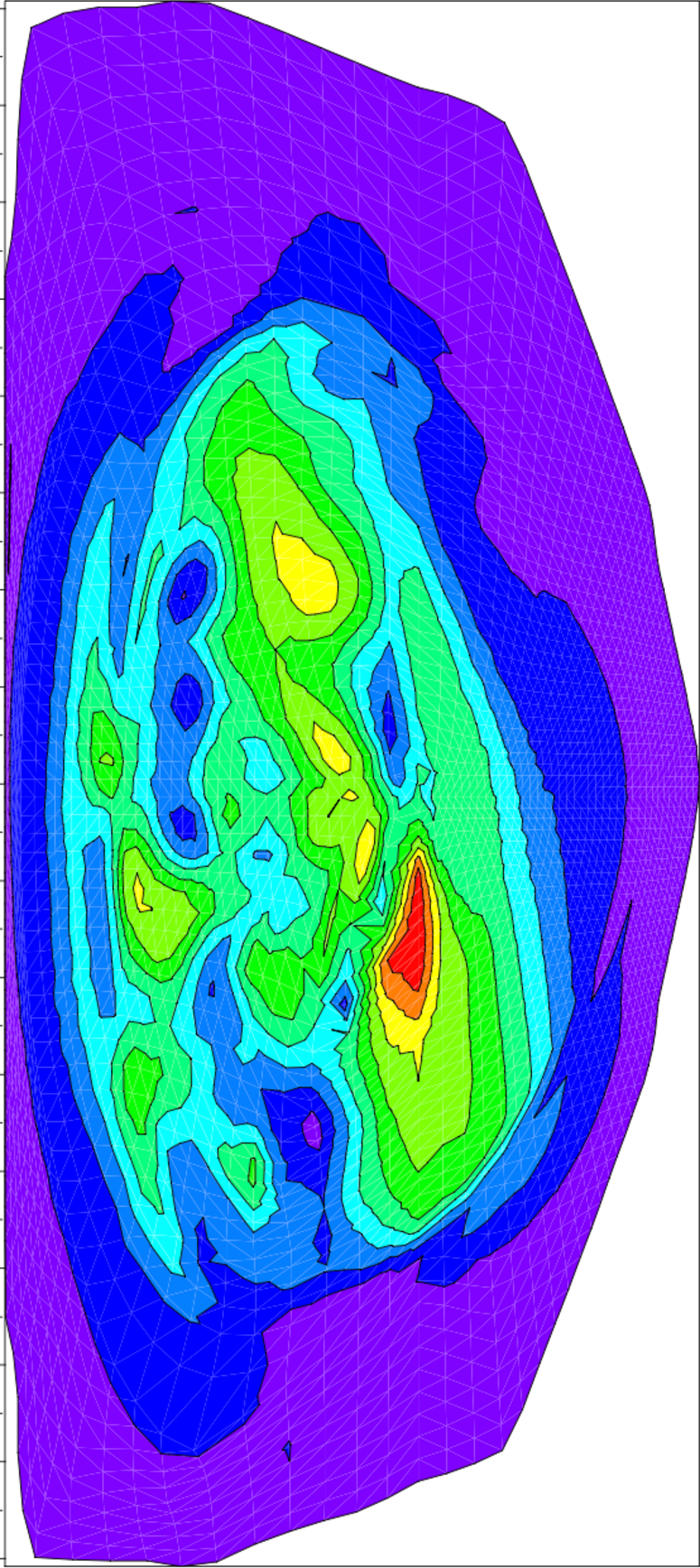}(c)
\end{center}
\caption{\it
Simulations with modified equilibrium reconstruction of  NSTX 
\cite{nstx25},
with $\beta_N = 3$, and resistive wall,  
(a) $\psi$ from feedback coils, compare with Fig. 2 (c); (b) pressure $p$ with feedback stabilization; (c)
$p$ without feedback, with (c) reproduced from \cite{nstx25} with AIP permission.} \label{fig:nstxfeed}
\end{figure}

Simulations were performed with a  modified equilibrium reconstruction of  NSTX example
\cite{nstx25},
with $\beta_N = 3$. \rfig{nstxfeed}(a) shows the flux $\psi$ produced by the coils.
The coils are  qualitatively similar to  the experiment \cite{sabbagh2010}.
This can be compared to \rfig{lobe} (c), indicating a reasonable overlap with the
unstable RWTM.
Contours of pressure  % for $\beta_N = 4$  %at $t \approx 800 \tau_A,$
  with feedback are shown in \rfig{nstxfeed}(b),
 while \rfig{nstxfeed} (c) 
shows  pressure contours without feedback, and a major disruption. 
%\end{section}

% \begin{section}{\Rdd{\bf   Current contraction criterion} } \label{sec:current} 
  \section*{\Rdd{\bf 6. Current contraction criterion} } \label{sec:current} 
The second disruption criterion in the DIII-D disruption database is the
critical $l_i / q_{95}.$
During locked mode disruption precursors the plasma can develop
low temperature in the edge. %Plasma is L mode.
This causes the current to contract.
%``The locked mode     usually exists for a long time before the
%disruption occurs, suggesting that  it is not always a primary cause of
%disruptions but an indicator of an unhealthy plasma condition." \cite{gerasimov2020}.
This has been  called a
``deficient edge" \cite{schuller}. % or
%``minor disruption" \cite{wesson}. 
Edge cooling in DIII-D has been produced by 
%$T_{e,q2}$ 
minor edge  disruptions  \cite{sweeney}.
Resistive ballooning turbulence can cause edge cooling, which might cause  
a density limit
\cite{ricci}. %or
%MARFE formation \cite{lipschultz1984}.  %and  resistive ballooning
The current contraction causes  increase of internal inductance $l_i$
as well as $\rho_{q2}.$
\begin{figure}[h]
\begin{center}
\includegraphics[height=5.2cm] {\figdir/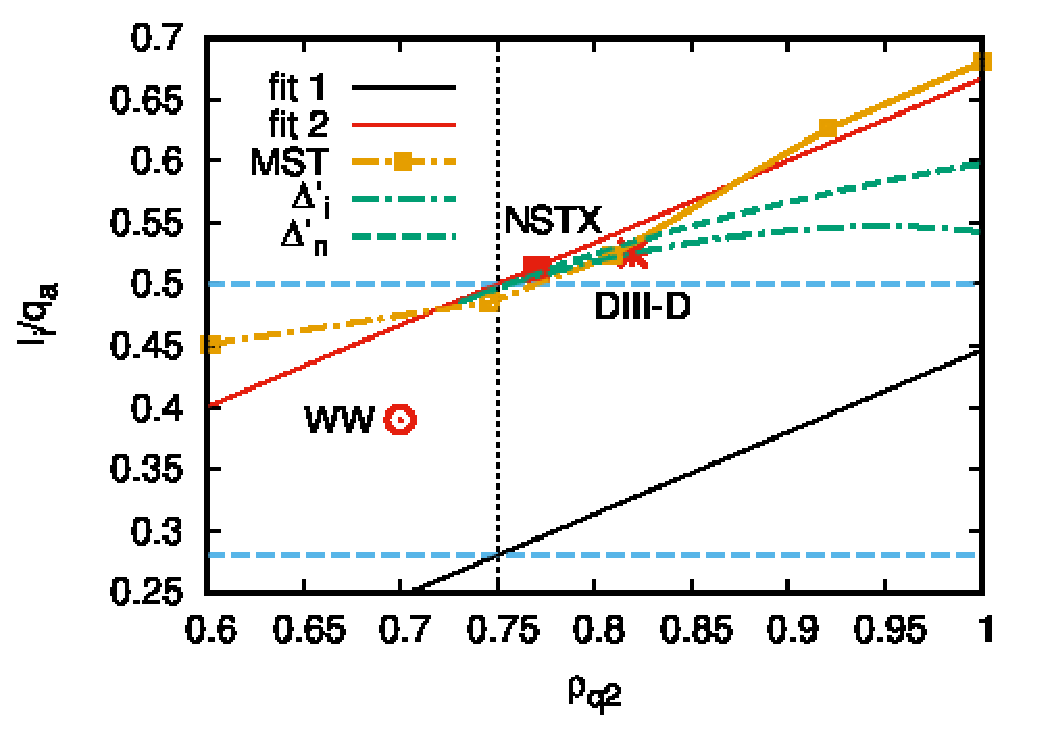}%(a) 
\end{center}
\caption{\it  $l_i/q_{95}$ as a function of $\rho_{q2}$. The data of \rfig{sweeney} is fit by ``fit 1."
It is shifted to agree \req{scale} with cicular boundary in ``fit 2". %The curve ``fit 3" is from
%the step current model \req{cheng}.
The connected points ``MST" are from \rfig{mst}. The curves $\Delta'_n, \Delta'_i$ are
from marginally stable analytic equilibria \cite{model}, 
with ideal wall and no wall.
They coalesce at $\rho_{q2} =
3/4, l_i / q_a = 1/2,$ when the boundary conditions are irrelevant. 
The point ``DIII-D" is from \cite{d3d22}, ``NSTX" from the simulations of \rfig{nstxfeed}, 
and ``WW" are 
simulations \cite{white80,waddell} with highly unstable initial conditions.} 
\label{fig:liq95}
\end{figure}
% \vspace{-.5cm}

\rfig{liq95} indicates how RWTMs can account for \rfig{sweeney}.
The line labelled ``fit 1" is \req{fit}, which fits the DIII-D data of \rfig{sweeney}. 
The circular boundary simulations labelled  
``MST"   are the  high $q_a$  MST \cite{model} simulations  in  \rfig{mst}. % \rsec{mst},
%The minor disruptions with $q_{q2} < 0.75, l_i / q_a < 0.5$ are connected with a dashed line.
% well fit by ``fit 2", $l_i / q_a = .67 \rho_{q2}.$   
The major  disruptions with $q_{q2} > 0.75, l_i / q_a > 0.5$
 align well with \req{scale} plotted as  ``fit 2." 
%$l_i / q_a = .67 \rho_{q2}.$   

The  critical values of $l_i / q_{95}$ for  $\rho_{q2} = 0.75$  vary as
$ l_i / q_{95} \approx 1/(2 \kappa),$ 
where $\kappa$ is the elongation, with
 $\kappa = 1$ in MST, and $\kappa \approx 1.76$ in  DIII-D. This onset scaling  also fits
 NSTX, with $\kappa \approx 2.5.$ 
This suggests combining   
the fits as 
%These two fits can be combined in ``fit 2", 
%given by
%\be \frac{l_i }{q_{95}} \stackrel{>}{\sim} \frac{1}{2 \kappa} \label{eq:scale} \ee
\be \frac{l_i }{q_{95}}  = \frac{2}{3}\rho_{q2}  -
\frac{1}{2} \left(1 - \frac{1}{\kappa} \right) \label{eq:scale} \ee
%where $\kappa$ is the elongation, and 
%in MST, $\kappa = 1$, and in  DIII-D, $\kappa \approx 1.76.$ This also fits  
% NSTX, with $\kappa \approx 2.5.$ 
which for $\rho_{q2} = 3/4$ recovers the onset scaling.
This allows geometry independent comparisons.
Note that only one onset condition is sufficient.
 The $\rho_{q2}$ onset condition determines the $l_i / q_{95} $  condition, and
{\it vice versa.}

\comment{
The curve ``fit 3" is the  step current model \cite{finn95,cheng},
\be \frac{l_i }{q_{a}} =\frac{1}{q_a}\left[\frac{1}{2} +\ln\left(\frac{q_a}{q_0}\right)\right] 
%= \frac{\rho_{q2}^2}{2}\left[ \frac{1}{2} 
%- \ln \left(\frac{\rho_{q2}^2}{2q_0}\right) \right] 
\label{eq:cheng} \ee 
with $q_0 = 1,$  
where $q_a = 2 / \rho_{q2}^2.$ In this model, the current is constant up to a
critical radius $(q_0/2)^{1/2} \rho_{q2}$  and zero for larger radius. 
% The curve ``fit 3" has $q_0 = 1.086.$ It corresponds to a marginally stable equilibrium
%\cite{d3d22,finn95}
%for $\rho_{q2} = 0.75,$ $\rho_w = 1.2.$  
For $\rho_{q2} < 0.75,$ it coincides with
``fit 2". For $\rho_{q2} > 0.75,$ it aligns closely with the curve $\Delta'_n.$
}

%In \rfig{liq95},
%``MST"  are  high $q_a$  MST \cite{model} simulations  in  \rfig{mst}. % \rsec{mst},
%The minor disruptions with $q_{q2} < 0.75, l_i / q_a < 0.5$ are connected with a dashed line.
% They align well with \req{scale} with $\kappa = 1,$  for $\rho_{q2} > 0.75.$ 

The
dashed curves $\Delta'_i, \Delta'_n$ in \rfig{liq95} are marginally stable TMs with
ideal and no wall, for model equilibria \cite{model,nf-iaea24} with variable 
current peaking \cite{frs}   and  current restricted in 
radius,  in  circular cylindrical geometry $\kappa = 1$. 
The marginal curves merge at $\rho_{q2} = 0.75,$ $l_i / q_a = .5,$ where the
boundary condition no longer affects the mode.
 The model predicts RWTM onset
quite well. As mentioned in Section 2, it also predicts the dependence of the onset 
on wall distance $\rho_w.$
%The curve $\Delta'_n$ agrees closely with ``fit 3"  \req{cheng}, which in turn agrees with
%``fit 2" \req{scale} for $\rho_{q2} \le 0.75.$

\comment{
The curve ``fit 3" is the  step current model \cite{finn95,cheng},
\be \frac{l_i }{q_{a}} =\frac{1}{q_a}\left[\frac{1}{2} +\ln\left(\frac{q_a}{q_0}\right)\right]
%= \frac{\rho_{q2}^2}{2}\left[ \frac{1}{2} 
%- \ln \left(\frac{\rho_{q2}^2}{2q_0}\right) \right]
\label{eq:cheng} \ee 
with $q_0 = 1,$  
where $q_a = 2 / \rho_{q2}^2.$ In this model, the current is constant up to a
critical radius $(q_0/2)^{1/2} \rho_{q2}$  and zero for larger radius.
% The curve ``fit 3" has $q_0 = 1.086.$ It corresponds to a marginally stable equilibrium
%\cite{d3d22,finn95}
%for $\rho_{q2} = 0.75,$ $\rho_w = 1.2.$
For $\rho_{q2} < 0.75,$ it coincides with
``fit 2". For $\rho_{q2} > 0.75,$ it aligns closely with the curve $\Delta'_n.$
}

The curves  ``fit 1" and ``fit 2" % and ``fit 3" 
%peaked  
correspond to equilibria which have
monotonic current
profiles with $q_0 = 1,$  and  zero current at $\rho =1,$  
such as the MST sequence in \rfig{mst}.
They are  consistent with   most of  the sampled cases in \rfig{sweeney}.
%}

In \rfig{liq95}
``DIII-D" is  shot 154576  \cite{d3d22}, and
``NSTX" is the  example in \rfig{nstxfeed},
where in both cases $l_i / q_{95} $ is increased \req{scale} by   $ .5(1 - 1/\kappa)$
to compare with ``fit 2".
%This  aligns  values from theory, simulation, and
%data independent of geometry. The agreement with the $\Delta_i', \Delta_n'$ curves,
%and 
This gives good agreement of  
the MST, DIII-D and NSTX  simulations.
%suggests that the disruptive DIII-D and NSTX modes are RWTMs.

%The  critical values of of $l_i / q_{95}$ for  $\rho_{q2} = 0.75$  vary as
%$ l_i / q_{95} \approx 1/(2 \kappa).$ 
%It should be noted that \req{fit},\req{scale} are for wall distance $\rho_w = 1.2.$
%The scalings might be generalized to take the wall distance into account.

The label
``WW" in \rfig{liq95} refers to  tearing mode  disruption simulations with 
highly unstable initialization \cite{white80,waddell}, $\kappa = 1,$
and  ideal wall. Qualitatively they explain disruptions, but they  do not fit the data.
%Disruptions can  be caused in simulations  if the plasma is initialized in a highly
%unstable initial state \cite{waddell, white80}. 
In other examples of highly unstable simulations, 
massive gas (MGI) or shattered pellet injection can cause disruptions by
quickly forcing the plasma into a highly unstable state \cite{izzo, nardon}.
In the case of locked modes, the plasma has hundreds of milliseconds to relax,
so a highly unstable state does not develop.

\comment{
\begin{figure}[h]
\begin{center}
 \includegraphics[height=5.2cm] {\figdirw/tmstable_qbq8-9-23c6c4g7e.eps}
\end{center}
\caption{\it $l_i(\rho_{q2})$. The instability criterion is $l_i > l_{i c}.$}
\label{fig:li}
\end{figure}
\rfig{liq95}(b) shows $l_i$ rather than $l_i / q_{95}$ as a function of $\rho_{q2}.$
The curves ``MST," ``$\Delta'_i$",  ``$\Delta'_n$", and ``fit 3" in \rfig{liq95} 
are plotted. The curve ``fit 4" is ``fit 2" in \rfig{liq95}(a),
 $l_i = 4/(3\rho_{q2}),$ where the step current model  is used to express $q_a = 2/\rho_{q2}^2.$ 
At the onset condition $\rho_{q2} = 3/4,$ $l_{i c} = 1 7/9.$
The curve ``$l_{i c}$"  is the instability boundary 
$l_{ic} = .5 q_a = \rho_{q2}^{-2},$ using the step current model. 
The figure also includes ``DIII-D",
``NSTX", and ``WW" points from \rfig{liq95}(a). 
Note that the instability condition $l_i > l_{i c}$ also includes $\rho_{q2} \ge 3/4.$
 The two onset  criteria can be combined into a
single criterion. 
}
In ITER baseline scenario
(IBS) equilibria in DIII-D, the instability condition  is more subtle than
a simple $l_i/q_{95}$ threshold \cite{turco}.
It depends on the details of the edge current profile,
in particular a local minimum of the edge current  or ``well" near $\rho_{q2} \approx 0.8.$
%and a local maximum  pedestal at $\rho \approx 0.9.$ 
% The value of $\Delta'$ increases with the well depth and pedestal height. 
 The value of $\Delta'$ can be large for a deep current well.
A way was found to produce the equilibria so they were relatively stable, by making a
shallower well and  perhaps decreasing  $\rho_{q2}$ so it was out of the well
and below the threshold.
%As mentioned,
%because the plasma has a long time to relax, it will not become highly unstable, and
%the well and pedestal should be small.
%The IBS equilibria satisfy $\rho_{q2} \approx 0.8,$
%and have $l_i / q_{95} \stackrel{>}{\sim} .28.$ 
The DIII-D example \cite{d3d22} 
 lacks the edge current structure. It should be pointed out that the database 
\cite{sweeney2017}  in \rfig{sweeney}  includes
some shots which do not disrupt but  satisfy the two criteria, 
as well as shots which disrupt without satisfying the criteria.
%The IBS equilibria satisfy $\rho_{q2} \approx 0.8.$ 
%When they disrupt, the unstable modes  are  RWTMs, because $\rho_{q2}$ is above the
%threshold for wall interaction.
%This  implies that IBS disruptions  can be feedback stabilized.

It has been shown that locked mode and high $\beta$ disruptions can be
caused by RWTMs. It has been commonly supposed that 
disruptions are caused by neoclassical tearing modes, %They require
but this is probably not the case.
Edge cooling suppresses edge current, including bootstrap current which drives NTMs. In simulations 
\cite{lahaye} NTMs
do not grow large enough for a major disruption. Typically they cause minor
disruptions and degrade confinement.
They are not important in IBS equilibria \cite{turco}.

It is noteworthy that ITER has RMP coils, which might be used for active feedback. This
might be possible because RWTMs are especially slow growing \cite{iter21} in ITER.
It would be highly desirable to prevent ITER disruptions, avoiding the need for
impurity injection and runaway electron generation. It should be relatively easy to
forecast RWTMs and rely  on the RMP coils instead of impurity injection.

An interesting prediction of \req{scale} is that the critical $l_i / q_{95}$ might
be controlled by varying the elongation. If an elliptical equilibrium is 
close to the onset condition, it might be possible to lower the elongation with
poloidal field coils and raise the critical $l_i / q_{95}$, preventing a disruption.
This is would be an $n = 0$ version of feedback stabilization.

%\end{section}
  
% \begin{section} {\Rdd{\bf Summary  }}
 \section*{\Rdd{\bf 7.  Summary  }}
There are two main criteria  for locked mode disruptions in DIII-D, 
which are signatures of RWTMs.
The first and more important, is that $q =2 $ rational surface radius 
normalized to plasma radius, $\rho_{q2},$
exceeds a threshold, which in DIII-D is given by
  $\rho_{q2} > 0.75.$ 
This is a  condition for a  RWTM. It  makes active feedback control possible.
Feedback control was demonstrated in experiments such as NSTX and KSTAR, and
in simulations of MST and NSTX.
The $\rho_{q2}$ depends on the normalized wall distance from the rational surface.

The second condition is that the internal inductance $l_i$ exceed a threshold,
which was expressed \cite{sweeney2017} in the  DIII-D database  as
   $li/q_{95} > 0.28.$ The internal inductance criterion is a condition for 
sufficient current peaking. 
Current peaking can be caused by edge cooling, which suppresses edge current and
bootstrap current. This includes suppression of edge bootstrap current, which 
 rules out NTMs as a cause of major disruptions.
The $l_i$ criterion depends on the plasma elongation.
 The dependence on  elongation is given, which  might be used as an actuator
 to prevent disruptions.
%Some shots do not disrupt but  satisfy the two criteria,
%while others disrupt without satisfying them. These cases should be
%investigated.}

   Simulations demonstrated that 
   RWTMs can grow to large amplitude and cause a complete
thermal quench. Ideal wall tearing modes grow to much smaller amplitudes, 
causing only minor disruptions, unless the plasma is 
initialized in a highly unstable state, which would not happen for locked modes. 
Feedback changes the resistive wall boundary condition  to effectively ideal.

RWTMs are found at high $\beta$ in NSTX and KSTAR, and  can be feedback stabilized.
Feedback stabilization of RWMs also stabilizes RWTMs.

 ITER resonant magnetic perturbation (RMP) coils
 might be used for feedback.

%\small
%Reference: H. R. Strauss, %Prevention of resistive wall tearing mode major disruptions with feedback,
%Phys. Plasmas \textbf{32} 032505 (2025); doi: 10.1063/5.0250999

%\end{section}
%{\bf Acknowledgement} 
\centerline{\small\bf ACKNOWLEDGEMENT}

This work was supported by  U.S. D.O.E. grant DE-SC0020127.

\small
%\input{bib1}
%\small

\end{document}